\documentclass[12pt,preprint]{aastex}  

\usepackage{pdfsync,epstopdf} %
\usepackage[dvipsnames,usenames]{color} %

\newcommand{\HII}{\mbox{H\thinspace{\sc ii}}} %

\shorttitle{Deuterium Fractionation in Massive Protostellar Cores} %
\shortauthors{Chen et al.} %

\begin{document} %
\title{
Deuterium Fractionation as an Evolutionary Probe in Massive Proto-stellar/cluster Cores
} %

\author{Huei-Ru Chen\altaffilmark{1,2}, Sheng-Yuan Liu\altaffilmark{2}, Yu-Nung Su\altaffilmark{2}, and Mei-Yan Wang\altaffilmark{1}} %

\altaffiltext{1}{Institute of Astronomy and Department of Physics, National Tsing Hua University, Hsinchu, Taiwan; hchen@phys.nthu.edu.tw.} %
\altaffiltext{2}{Institute of Astronomy and Astrophysics, Academia Sinica, Taipei, Taiwan.} %

\begin{abstract} %
Clouds of high infrared extinction are promising sites of massive star/cluster formation.
A large number of cloud cores discovered in recent years allows investigation of possible evolutionary sequence among cores in early phases.      
We have conducted a survey of deuterium fractionation toward 15 dense cores in various evolutionary stages, from high-mass starless cores to ultracompact \HII\ regions, in the massive star-forming clouds of high extinction, G34.43+0.24, IRAS 18151$-$1208, and IRAS 18223-1243, with the Submillimeter Telescope (SMT).  
Spectra of $\mathrm{N_2H^+} \; (3-2)$, $\mathrm{N_2D^+} \; (3-2)$, and $\mathrm{C^{18}O} \; (2-1)$  were observed to derive the deuterium fractionation of $\mathrm{N_2H^+}$, $D_\mathrm{frac} \equiv N(\mathrm{N_2D^+})/N(\mathrm{N_2H^+})$, as well as the CO depletion factor for every selected core.   
Our results show a decreasing trend in $D_\mathrm{frac}$ with both gas temperature and linewidth. 
Since colder and quiescent gas is likely to be associated with less evolved cores, larger $D_\mathrm{frac}$ appears to correlate with early phases of core evolution.
Such decreasing trend resembles the behavior of $D_\mathrm{frac}$ in the low-mass protostellar cores and is consistent with several earlier studies in high-mass protostellar cores.  
We also find a moderate increasing trend of $D_\mathrm{frac}$ with the CO depletion factor, suggesting that sublimation of ice mantles alters the competition in the chemical reactions and reduces $D_\mathrm{frac}$. 
Our findings suggest a general chemical behavior of deuterated species in both low- and high-mass proto-stellar candidates at early stages.  
In addition, upper limits to the ionization degree are estimated to be within $2 \times 10^{-7}$ and $5 \times 10^{-6}$.  
The four quiescent cores have marginal field-neutral coupling and perhaps favor turbulent cooling flows. 
\end{abstract} %


\section{Introduction} %
Clouds of high infrared (IR) extinction, such as infrared dark clouds (IRDCs), are discovered through silhouette against the bright, diffuse IR background emission of the Galactic plane \markcite{Egan1998ApJ494,Rathborne2006ApJ641,Rygl:2010bq}(Egan {et~al.} 1998; Rathborne, Jackson, \&  Simon 2006; Rygl {et~al.} 2010).
Because of their cold ($T \lesssim 20 \; \mathrm{K}$), dense ($n_\mathrm{H_2} \gtrsim 10^4 \; \mathrm{cm^{-3}}$), and massive ($M \gtrsim 10^3  M_\odot$) nature, IRDCs are thought to be promising sites of massive star/cluster formation.  
They often harbor cores in a sequence of evolutionary stages \markcite{Beuther2007ApJ668,Beuther2010A&A518}(Beuther \& Sridharan 2007; Beuther {et~al.} 2010), from fairly quiescent high-mass starless cores \markcite{Sridharan2005ApJ634,Beuther2007ApJ668}(HMSCs; Sridharan {et~al.} 2005; Beuther \& Sridharan 2007), followed by cores with accreting low/intermediate-mass proto-stars/clusters \markcite{Beuther2007ApJ656,Wang:2011jl}(Beuther \& Steinacker 2007; Wang {et~al.} 2011), to high-mass protostellar objects \markcite{Sridharan2002ApJ566,Beuther2002ApJ566}(HMPOs; Sridharan {et~al.} 2002; Beuther {et~al.} 2002a), and even to ultra-compact \HII\ (UC \HII) regions \markcite{Battersby2010ApJ721}(Battersby {et~al.} 2010).  
In recent years, a large number of massive cloud cores ($M \gtrsim 10^2  M_\odot$) have been identified in IRDCs \markcite{Peretto:2010kb}(Peretto \& Fuller 2010) and enable the investigation of possible evolutionary sequence among cores in early phases \markcite{Chambers2009ApJS181,Battersby2010ApJ721,Henning2010A&A518}(Chambers {et~al.} 2009; Battersby {et~al.} 2010; Henning {et~al.} 2010).    
In general, the strategy to probe evolutionary stages usually involves several indicators, each of which traces some star-forming activities, e.g. radio emission for ionizaiton, gas or dust temperature for internal heating, molecular linewidth for turbulence, asymmetry of line profiles for infall motions, or high-velocity gas for outflows \markcite{Crapsi2005ApJ619,Battersby2010ApJ721}(e.g. Crapsi {et~al.} 2005; Battersby {et~al.} 2010).   
In our previous study of G28.34+0.06, we found a moderate decreasing trend in the deuterium fractionation of $\mathrm{N_2H^+}$, $D_\mathrm{frac} \equiv N(\mathrm{N_2D^+})/N(\mathrm{N_2H^+})$, with evolutionary stage in three selected cores \markcite{Chen2010ApJ713}(Chen {et~al.} 2010).  
Here we further investigate the possibility to use $D_\mathrm{frac}$ as an evolutionary probe for high-mass protostellar/cluster cores by searching its correlation with some indicators in three more clouds of high IR extinction.  

In early stages of star formation process, the low-temperature environment nurtures a peculiar chemistry with high deuterium enrichment as a result of exothermic deuteration reactions and significant depletion of CO and other neutral species.  
Deuterium enrichment is primarily initiated by the reaction
\begin{equation} %
\mathrm{H_3^+ + HD \leftrightharpoons H_2D^+ + H_2,}  \label{eq_HD}  
\end{equation} %
which is exothermic in the forward direction with $\Delta E/k = 230 \; \mathrm{K}$ \markcite{Millar1989ApJ340}(Millar, Bennett, \& Herbst 1989).  
At very low temperature ($T < 20 \; \mathrm{K}$), the back reaction becomes negligible, which results in an enhancement of $\mathrm{H_2D^+}$ \markcite{Stark1999ApJ521,Stark2004ApJ608,Caselli2003A&A403,Caselli:2008gk,Vastel2006ApJ645,vanderTak2005A&A439,Harju:2006dj,Friesen2010ApJ718}(Stark, van~der Tak, \& van  Dishoeck 1999; Stark {et~al.} 2004; Caselli {et~al.} 2003, 2008; Vastel {et~al.} 2006; van~der Tak, Caselli, \&  Ceccarelli 2005; Harju {et~al.} 2006; Friesen {et~al.} 2010) and even multiply deuterated $\mathrm{H_3^+}$ \markcite{Vastel2004ApJ606,Parise:2010fk,Roberts2003ApJ591}(Vastel, Phillips, \& Yoshida 2004; Parise {et~al.} 2010; Roberts, Herbst, \& Millar 2003).  
Deuteration is further enhanced in cold cores after the removal of the gas-phase CO, the major destroyer of $\mathrm{H_2D^+}$, due to depletion of molecular species on dust grains \markcite{Ceccarelli2007PPV,Bergin2007ARA&A45,vanderTak:2006ip,Aikawa2008Ap&SS313,Millar:2005ce}(Ceccarelli {et~al.} 2007; Bergin \& Tafalla 2007; van~der Tak 2006; Aikawa 2008; Millar 2005).  
Indeed, deuterated species are often observed with an enhancement of $2-3$ orders of magnitude in star-forming cores \markcite{Crapsi2005ApJ619,Fontani2006A&A460,Fontani2011A&A529,Pillai2007A&A467}(Crapsi {et~al.} 2005; Fontani {et~al.} 2006, 2011; Pillai {et~al.} 2007) over the local interstellar value of $1.51 \times 10^{-5}$ \markcite{Oliveira2003ApJ587}(Oliveira {et~al.} 2003).  
Chemical models anticipate some correlation between the deuterium fractionation and the CO depletion factor, $f_D$, which increases in the prestellar phase and declines later in the protostellar phase as the mantle sublimation occurs \markcite{Ceccarelli2007PPV,Crapsi2005ApJ619,Emprechtinger2009A&A493}(Ceccarelli {et~al.} 2007; Crapsi {et~al.} 2005; Emprechtinger {et~al.} 2009).
Such correlation has been recognized with the deuterium fractionation of $\mathrm{N_2H^+}$ in a complied sample of low-mass prestellar and protostellar cores, among which cores in same molecular clouds show better correlation that indicates environmental influences, such as magnetic field strength, amount of turbulence, external radiation field, etc \markcite{Crapsi2005ApJ619,Emprechtinger2009A&A493}(Crapsi {et~al.} 2005; Emprechtinger {et~al.} 2009).  
On the other hand, early studies of high-mass protostellar objects often did not show a consistent behavior of deuterium fractionation with evolutionary stage other than a general enhancement \markcite{Fontani2006A&A460,Pillai2007A&A467}(Fontani {et~al.} 2006; Pillai {et~al.} 2007).  
In the study of G28.34+0.06  \markcite{Chen2010ApJ713}(Chen {et~al.} 2010), we started off our sample with cores in one IRDC and found a moderate decreasing trend in the deuterium fractionation of $\mathrm{N_2H^+}$ with evolutionary stages from a massive starless core (MM9), a more evolved core with fragmentation and outflows \markcite{Wang:2011jl}(MM4; Wang {et~al.} 2011), to an UC \HII\ region (MM1).  
Subsequently, \markcite{Fontani2011A&A529}Fontani {et~al.} (2011) significantly improved the statistics with a larger sample of twenty-seven cores and also revealed a similar trend across evolutionary phases from HMSCs, HMPOs, to UC \HII\ regions.  
In particular, the authors reported a nice decreasing trend with gas temperature and column density of $\mathrm{N_2H^+}$.
Recently, \markcite{Miettinen:2011um}Miettinen, Hennemann, \&  Linz (2011) reported a similar decreasing trend but in the $N(\mathrm{DCO^+})/N(\mathrm{HCO^+})$ ratio with gas temperature. 
They further made first estimates of ionization fraction and cosmic ionization rate in massive IRDCs.  
Over all, the use of the $N(\mathrm{N_2D^+})/N(\mathrm{N_2H^+})$ ratio as an evolutionary probe to high-mass protostellar candidates has been suggested \markcite{Chen2010ApJ713,Fontani2011A&A529,Caselli:2011tb,Miettinen:2011um}(Chen {et~al.} 2010; Fontani {et~al.} 2011; Caselli 2011; Miettinen {et~al.} 2011).    

We select cores in three massive star-forming regions of similar luminosity ($L \sim 10^4 \; L_\odot$) to probe a sequence of evolutionary stages: one IRDC, G34.43+0.24 \markcite{Rathborne2006ApJ641}(hereafter G34.43; Rathborne {et~al.} 2006), and two HMPOs, IRAS~18151$-$1208 and IRAS~18223$-$1243 \markcite{Sridharan2002ApJ566}(hereafter I18151 and I18223, respectively; Sridharan {et~al.} 2002).  
All three regions harbor multiple cores at different evolutionary stages, and discussions of individual clouds are provided in the Appendix.  
To better parameterize the evolutionary stage with gas properties, we limit our sample to cores with available ammonia gas temperatures from a survey conducted by \markcite{Sakai2008ApJ678}Sakai {et~al.} (2008).  
This renders eight cores in G34.43, three cores in I18151, and four cores in I18223.  
Table~\ref{t1} lists the selected cores and their known properties such as gas temperature, $T_g$, integrated flux density at $1.2 \; \mathrm{mm}$, $F_\mathrm{1.2 \, mm}$, core size, $2 R$, molecular  number density, $n_\mathrm{H_2}$.  
In addition, four out of the fifteen selected cores are not associated any known outflows (Table~\ref{t1}) and possibly at the earliest stage.  
We consider them to be quiescent and perhaps pre-stellar cores in this study.  
Our sample of cores have gas temperatures from $14$ to $21 \; \mathrm{K}$, which are within a promising range to detect variation in $f_D$ based on dust mantle sublimation \markcite{Collings2003Ap&SS285}(Collings {et~al.} 2003). 
These cores are generally colder than those considered in previous studies \markcite{Fontani2006A&A460,Fontani2011A&A529}(Fontani {et~al.} 2006, 2011).
 
Based on previous studies, e.g. \markcite{Chen2010ApJ713}Chen {et~al.} (2010) and \markcite{Fontani2011A&A529}Fontani {et~al.} (2011), we would like to  assess the use of $D_\mathrm{frac}$ as an evolutionary probe with a large sample of cores, particularly for evolutionary stages before HMCs and UC \HII\ regions.  
In this study, a similar strategy is employed to compare $D_\mathrm{frac}$ among cores in similar IRDCs to reduce the environmental fluctuations among selected cores.  
Moreover, we also investigate the behavior of the $N(\mathrm{N_2D^+})/N(\mathrm{N_2H^+})$ ratio with the CO depletion factor through $\mathrm{C^{18}O}$ observations and estimate the electron abundance. 
In \S~\ref{sec_Observations}, we describe the observations and related parameters.    
We then explain in \S~\ref{sec_SpectralFits} the derivation of $D_\mathrm{frac}$ with self-consistent spectral fits for $\mathrm{N_2H^+}$ and $\mathrm{N_2D^+}$ as well as the determination of the CO depletion factor.  
Lastly, in \S~\ref{sec_Results}, we discuss the use of the $N(\mathrm{N_2H^+})/N(\mathrm{N_2D^+})$ ratio as an evolutionary probe as well as the upper limits for the degree of ionization and its implications.   



\section{Observations and Data Reduction \label{sec_Observations}} %
We observed $\mathrm{N_2H^+ \; (3-2)}$ at 279.511780~GHz, $\mathrm{N_2D^+ \; (3-2)}$ at 231.321864~GHz, and $\mathrm{C^{18}O \; (2-1)}$  at 219.56036~GHz towards selected cores in G34.43,  I18151, and I18223 with the 10-meter Arizona Radio Observatory (ARO) Submillimeter Telescope (SMT) on Mount Graham, Arizona.  
The observations were carried out between November 2008 and April 2011 in the beam-switching (BS) mode for $\mathrm{N_2H^+}$ and $\mathrm{N_2D^+}$ and in the absolute position-switching (APS) mode for $\mathrm{C^{18}O}$ to correctly subtract the background because $\mathrm{C^{18}O}$ emission is supposed to be extended.  
The pointing centers are given in Table~\ref{t1} with typical uncertainty of $\sim 3\arcsec$.   
The primary beam is about $27^{\prime\prime}$ for $\mathrm{N_2H^+}$, $32^{\prime\prime}$ for $\mathrm{N_2D^+}$, and $34^{\prime\prime}$ for $\mathrm{C^{18}O}$. 
The spectral resolution is $1 \; \mathrm{MHz}$, corresponding to a velocity resolution of $1.07$ and $1.30 \; \mathrm{km \, s^{-1}}$ for $\mathrm{N_2H^+}$ and $\mathrm{N_2D^+}$, respectively, and $0.25 \; \mathrm{MHz}$, equivalent to $0.34 \; \mathrm{km \, s^{-1}}$, for $\mathrm{C^{18}O}$. 
The temperature scale $T_\mathrm{A}^*$ was obtained using standard vane calibration, and the main beam temperature, $T_\mathrm{mb}$, was derived through $T_\mathrm{mb} = T_\mathrm{A}^*/\eta_\mathrm{mb}$ with a main beam efficiency $\eta_\mathrm{mb} = 0.75$. 
Typical system temperature during the observations was around $200 - 300 \; \mathrm{K}$ and the respective rms noise level for each observation is given in Table~\ref{t1}. 
Data reduction was performed with the CLASS package \markcite{Guilloteau:2000vr}(Guilloteau \& Lucas 2000, ; see also http://www.iram.fr/IRAMFR/GILDAS).  
 
\section{Data Analysis and Spectral Fits \label{sec_SpectralFits}} %
\begin{figure} %
\plotone{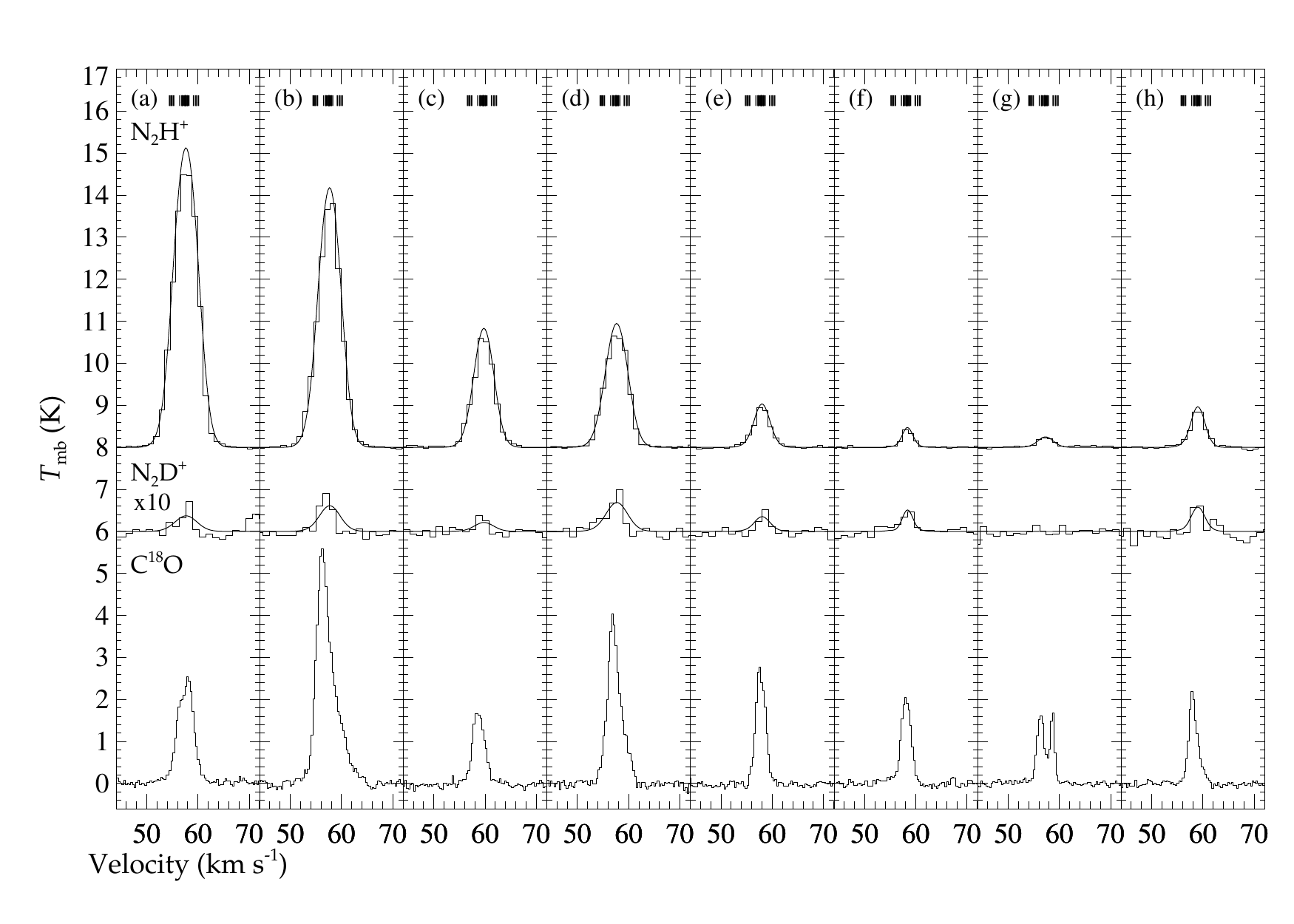} %
\caption{(a)--(h) Spectra of $\mathrm{N_2H^+ \; (3-2)}$ shifted by $8 \; \mathrm{K}$ (top histogram) in the order of G34.43--MM1, MM2, MM3, MM4, MM5, MM6, MM8, and MM9, superposed with the thirty-eight hyperfine component synthesis spectra (solid curves) assuming a single excitation temperature based on $\mathrm{NH_3}$ observations \markcite{Sakai2008ApJ678}(Sakai {et~al.} 2008).  
The frequency of each individual hyperfine component in the model is labelled with a short bar on the top.  
Spectra of $\mathrm{N_2D^+ \; (3-2)}$ multiplied by $10$ and shifted by $6 \; \mathrm{K}$ (middle histogram) toward the corresponding cores superposed with synthesis spectra (solid curves) and spectra of $\mathrm{C^{18}O \; (2-1)}$ (bottom histogram) are also shown.  
} %
\label{f1} %
\end{figure}

\begin{figure} %
\plotone{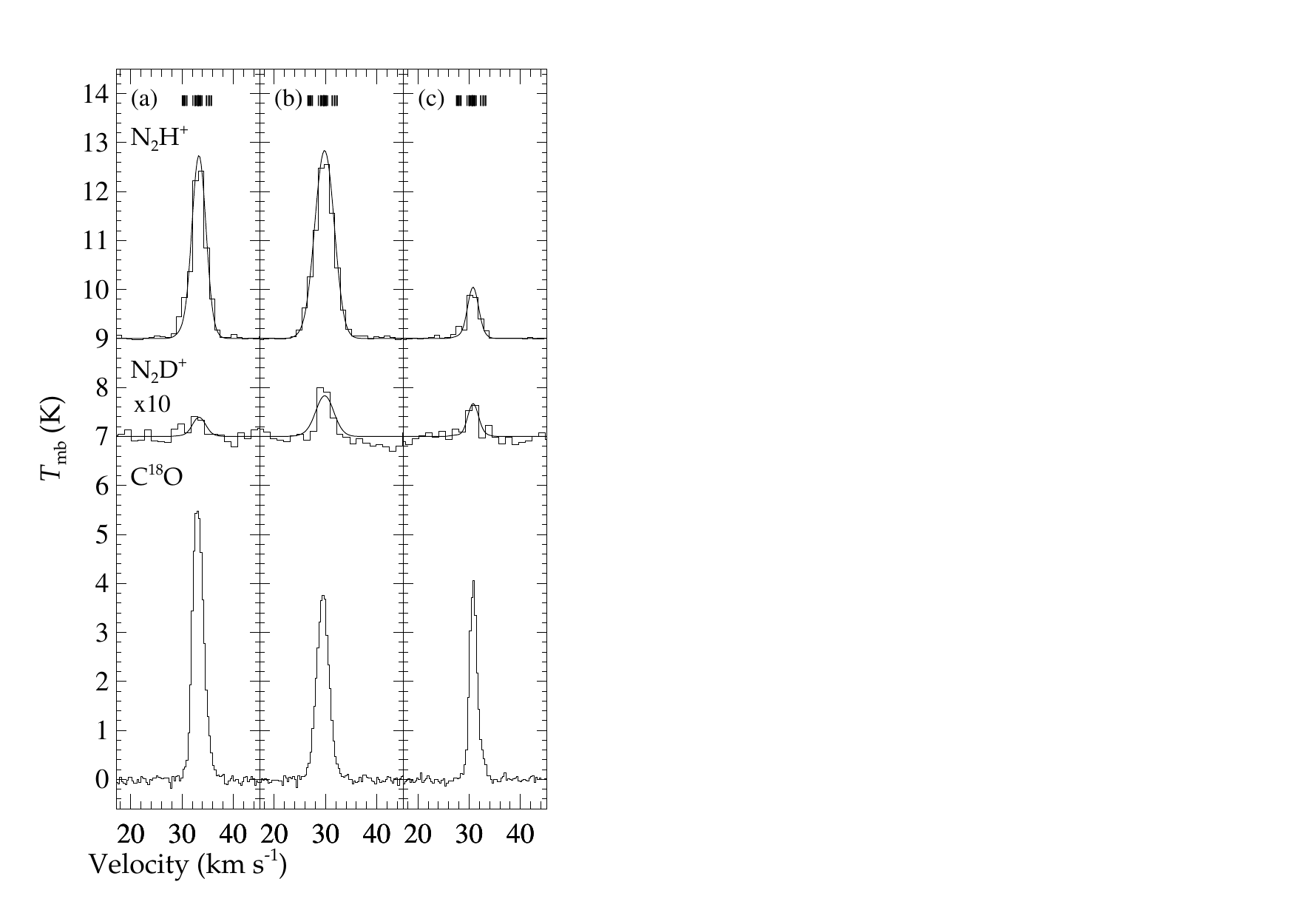} %
\caption{(a)--(c) Similar to Fig.~\ref{f1} but for I18151--MM1, MM2, and MM3. 
Spectra of $\mathrm{N_2H^+} \; (3-2)$ are shifted by $9 \; \mathrm{K}$ and those of $\mathrm{N_2D^+} \; (3-2)$ are multiplied by 10 and shifted by $7 \; \mathrm{K}$. 
} %
\label{f2} %
\end{figure}

\begin{figure} %
\plotone{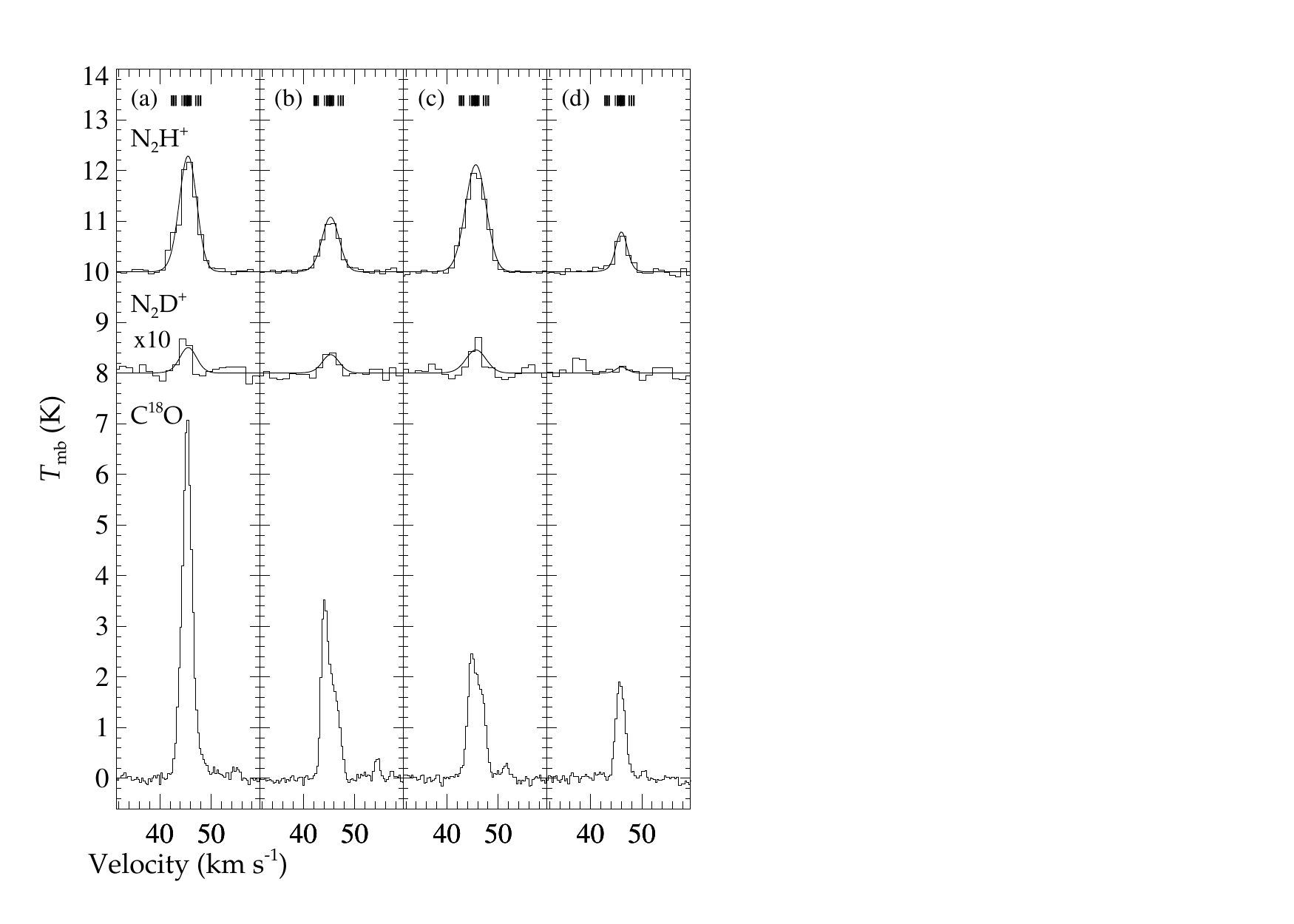} %
\caption{(a)--(d) Similar to Fig.~\ref{f1} but for I18223--MM1, MM2, MM3, and MM4.  
Spectra of $\mathrm{N_2H^+} \; (3-2)$ are shifted by $10 \; \mathrm{K}$ and those of $\mathrm{N_2D^+} \; (3-2)$ are multiplied by 10 and shifted by $8 \; \mathrm{K}$. 
} %
\label{f3} %
\end{figure}

Except the $\mathrm{N_2D^+}$ in G34.43--MM8 (Fig.~\ref{f1}g), all three molecular lines are detected in emission toward every core.  
The spectra for cores in G34.43 are shown in Fig.~\ref{f1} while those for in I18151 and I18223 in Fig.~\ref{f2} and \ref{f3}, respectively.  
We derive $D_\mathrm{frac}$ by analyzing the $\mathrm{N_2H^+}$ and $\mathrm{N_2D^+}$ spectra with a self-consistent spectral fit.   
The CO depletion factor, $f_D$, is also estimated with the observed $\mathrm{C^{18}O}$ spectra and available 1.2~mm continuum flux density.  

\subsection{Self-Consistent Spectral Fits for $\mathrm{N_2H^+}$ and $\mathrm{N_2D^+}$} %
In some cores, the emission of $\mathrm{N_2D^+}$ is undesirably weak and becomes more challenging to constrain the spectral fit parameters.  
Instead of performing an independent spectral fit for $\mathrm{N_2D^+}$, we improve the model described in \markcite{Chen2010ApJ713}Chen {et~al.} (2010) to perform a self-consistent fit for the $\mathrm{N_2H^+}$ and $\mathrm{N_2D^+}$ spectra using $D_\mathrm{frac}$ as a scaling factor.   
Because of the numerous hyperfine components, we fit each spectrum of every core with a synthetic  spectrum comprised of thirty-eight hyperfine components with updated line frequencies and spontaneous emission rates \markcite{Pagani2009A&A494}(Pagani, Daniel, \& Dubernet 2009). 
For each individual source, all the hyperfine components of every $J$-level are assumed to be in thermal equilibrium at a single excitation temperature, $T_g$, adopted from the ammonia gas temperature in \markcite{Sakai2008ApJ678}Sakai {et~al.} (2008).
The synthetic spectra are described by three more parameters: the total column density, $N(\mathrm{N_2H^+})$, the systemic velocity, $\upsilon_\mathrm{LSR}$, and the full-width at half maximum (FWHM) as line width, $\Delta \upsilon$.
Model spectra are optimized with the minimization of the reduced $\chi^2$ value, $\overline{\chi^2}$, and the results are listed in Table~\ref{t2}.  
Note that the uncertainties of the adopted $T_g$ \markcite{Sakai2008ApJ678}(Sakai {et~al.} 2008) are also incorporated into the uncertainties of all the derived spectral parameters.  

Taking the cosmic background temperature, $T_\mathrm{bg} = 2.7 \; \mathrm{K}$ into account, we  derive the optical depth of an observed spectrum with  
\begin{equation} %
  \tau(\upsilon) = -\ln \left[ 1 - \frac{T_\mathrm{mb}(\upsilon)}{J(T_g) - J(T_\mathrm{bg})} \right],  \label{eq_tau}
\end{equation} %
where $T_\mathrm{mb}(\upsilon)$ is the main beam temperature of the spectra, and $J(T) \equiv (h\nu/k)/(e^{h\nu/kT}-1)$.
In general, the spectra have fairly small optical depths if a beam filling factor of unity is assumed.  
The optimized spectral model also delivers an estimate for the $\mathrm{N_2H^+}$ optical depth by integrating optical depths of all the hyperfine components.
The maximum optical depth, $\tau_\mathrm{max}$, is about 0.8 in G34.43-MM1 (Table~\ref{t2}).
Emissions of $\mathrm{N_2H^+}$ and $\mathrm{N_2D^+}$ in all the observed cores remain optically thin.  
Although sub-structures within our observing beam sizes cannot be completely ruled out, the emission in these early phases of core evolution is mostly attributed to large scales.  
In the example of G28.34+0.06, the integrated flux observed with the Submillimeter Array (SMA) is less than 10\% of the total flux observed with the single-dish telescope SMT \markcite{Chen2010ApJ713}(Chen {et~al.} 2010).  

\subsection{The CO Depletion Factor \label{sec_fD}} %
The CO depletion factor, $f_D$, is defined as the ratio of the canonical CO abundance, $x(\mathrm{CO})_\mathrm{can}$, to the observed CO abundance, $x(\mathrm{CO})_\mathrm{obs}$,   
\begin{equation} %
f_D \equiv \frac{x(\mathrm{CO})_\mathrm{can}}{x(\mathrm{CO})_\mathrm{obs}} = \frac{x(\mathrm{C^{18}O})_\mathrm{can}}{x(\mathrm{C^{18}O})_\mathrm{obs}}.   
\label{eq_fD}
\end{equation} %
Using the $\mathrm{C^{18}O}$ abundance of $1.7 \times 10^{-7}$  in the solar neighborhood \markcite{Frerking1982ApJ262}(Frerking, Langer, \&  Wilson 1982) and the abundance gradients of $\Delta \log [\mathrm{C}/\mathrm{H}]/\Delta R = -0.066 \; \mathrm{dex \, kpc^{-1}}$ and $\Delta \log [\mathrm{O}/\mathrm{H}]/\Delta R = -0.065 \; \mathrm{dex \, kpc^{-1}}$ in the Galactic Disk \markcite{Wilson1992A&ARv4}(Wilson \& Matteucci 1992), the canonical abundance of $\mathrm{C^{18}O}$ is estimated to be 
\begin{equation} %
x(\mathrm{C^{18}O})_\mathrm{can} = 1.7 \times 10^{-7} \cdot 10^{-0.131 \, (D_\mathrm{GC}-D_\odot)}, 
\end{equation} %
where $D_\mathrm{GC}$ is the Galactocentric distance of the core, and $D_\odot = 8.5 \; \mathrm{kpc}$ is the distance of the Sun to the Galactic Center.      
Given the location of our IRDCs, $x(\mathrm{C^{18}O})_\mathrm{can} = 3.80 \times 10^{-7}$, $3.92 \times 10^{-7}$, and $4.68 \times 10^{-7}$ for G34.43, I18151, and I18223, respectively.    

Over all, the observed brightness temperature, $T_\mathrm{mb}$, is smaller than the kinetic temperature, $T_g$, and renders fairly small optical depths.  
The maximum optical depth estimated with Eq.~(\ref{eq_tau}) in individual IRDC is 0.5 for G34.43-MM2, 0.5 for I18151-MM3, and 0.8 for I18223-MM1.      
Assuming that all rotational levels are thermalized, we determine the column density of $\mathrm{C^{18}O}$ with the method based on \markcite{Caselli2002ApJ565b}Caselli {et~al.} (2002) that accommodates the effect of  background emission at $T_\mathrm{bg}$,  
\begin{equation} %
N(\mathrm{C^{18}O}) = \frac{3 h}{8 \pi^3} \frac{1}{\mu^2 S_{21}} \, W(\mathrm{C^{18}O}) \, \frac{Q(T_g)}{T_g - T_\mathrm{bg}} \, \frac{e^{E_\mathrm{up}/k T_g}}{(e^{h\nu_0/k T_g} -1)}, 
\end{equation} %
where $Q(T_g)$ is the partition function, $\mu^2 \, S_{21} = 0.02440 \; \mathrm{Debye}^2$ for the $J=2-1$ transition, $\nu_0$ the transition rest frequency, $E_\mathrm{up} = 15.8 \; \mathrm{K}$ the upper level energy, and $W(\mathrm{C^{18}O})$ the integrated brightness temperature in velocity  (Table~\ref{t1}).   
Since some of the spectra do not resemble a Gaussian profile, direct integration in the channels with significant emission is performed to obtain $W(\mathrm{C^{18}O})$ without fitting a Gaussian profile, and the derived $N(\mathrm{C^{18}O})$ is listed in Table~\ref{t3}.     

The observed $\mathrm{C^{18}O}$ fractional abundance, $x(\mathrm{C^{18}O})_\mathrm{obs}$,  depends on the column density of $\mathrm{H_2}$, $N_\mathrm{H_2}$, and $x(\mathrm{C^{18}O})_\mathrm{obs} = N(\mathrm{C^{18}O})_\mathrm{obs}/N_\mathrm{H_2}$.  
In the attempt to reduce the uncertainty in deriving $x(\mathrm{C^{18}O})_\mathrm{obs}$, we first match the angular resolutions between the $\mathrm{C^{18}O}$ and dust continuum observations by  convolving the 1.2~mm continuum maps in the literature \markcite{Beuther2002ApJ566,Rathborne2006ApJ641}(beam size = 11\arcsec; Beuther {et~al.} 2002a; Rathborne {et~al.} 2006) with the $34^{\prime\prime}$ beam of our $\mathrm{C^{18}O}$ observations.
The column density of $\mathrm{H_2}$ is estimated with the 1.2~mm peak flux density, $S_\mathrm{1.2 \, mm}$, arising from warm dust 
\begin{equation} %
N_\mathrm{H_2} = \frac{S_\mathrm{1.2 \, mm}}{\kappa_\mathrm{1.2 \, mm} \, B_\nu(T_d) \, \Omega_b} \frac{1}{\mu \, m_\mathrm{H_2}}, 
\end{equation} %
where $\kappa_\mathrm{1.2 \, mm} = 0.005 \; \mathrm{cm^2 \, g^{-1}}$ is the dust opacity assuming a gas-to-dust mass ratio of $100$ \markcite{Shepherd2002ApJ566}(Shepherd \& Watson 2002), $B_\nu(T_d)$ the Planck function at dust temperature, $T_d$, $\Omega_b$ the solid angle subtended by the convolved beam size of $34^{\prime\prime}$, $\mu = 1.36$ the mean molecular weight, and $m_\mathrm{H_2}$ the mass of $\mathrm{H_2}$ molecule. 

For warmer cores in our sample, the averaged dust temperature, $T_d$, have been found from studies of the spectral energy distributions (SEDs) \markcite{Rathborne2005ApJ630,Marseille2008A&A488,Beuther2010A&A518}(Table~3; Rathborne {et~al.} 2005; Marseille {et~al.} 2008; Beuther {et~al.} 2010).  
However, most cores in our sample remain as extinction features in the near- or mid-IR, and it is  challenging to derive their dust temperatures.   
Sensitive mid- and far-IR observations, such as {\it Herschel}, will offer better opportunities to  constrain dust emission properties, including temperature, in cold clumps of IRDCs \markcite{Beuther2010A&A518,Stutz2010A&A518}(e.g. Beuther {et~al.} 2010; Stutz {et~al.} 2010).  
Alternatively, we adopt gas temperature, $T_g$, for the conversion between $S_\mathrm{1.2 \, mm}$ to $N_\mathrm{H_2}$ when dust temperature is unavailable.  
In case the thermal coupling between dust and gas is not ideal, this assumption may underestimate $T_d$  and hence overestimate $N(\mathrm{H_2})$ by a factor of $< 2$ in our sample. 
In the case of G34.43--MM2, no SED study is found to constrain its dust temperature.  
Since the core is associated with an UC \HII\ region, we expect its averaged dust temperature to be warmer with respect to MM1 and sets a lower limit of $T_d \ge 34 \; \mathrm{K}$ for MM2.  

Once the observed $\mathrm{C^{18}O}$ abundance, $x(\mathrm{C^{18}O})_\mathrm{obs}$, is determined, the CO depletion factor, $f_D$, can be computed with Eq.~(\ref{eq_fD}) accordingly, and   
the results are summarized in Table~\ref{t3}.  

\section{Results and Discussion \label{sec_Results}} %

\begin{figure} %
\plotone{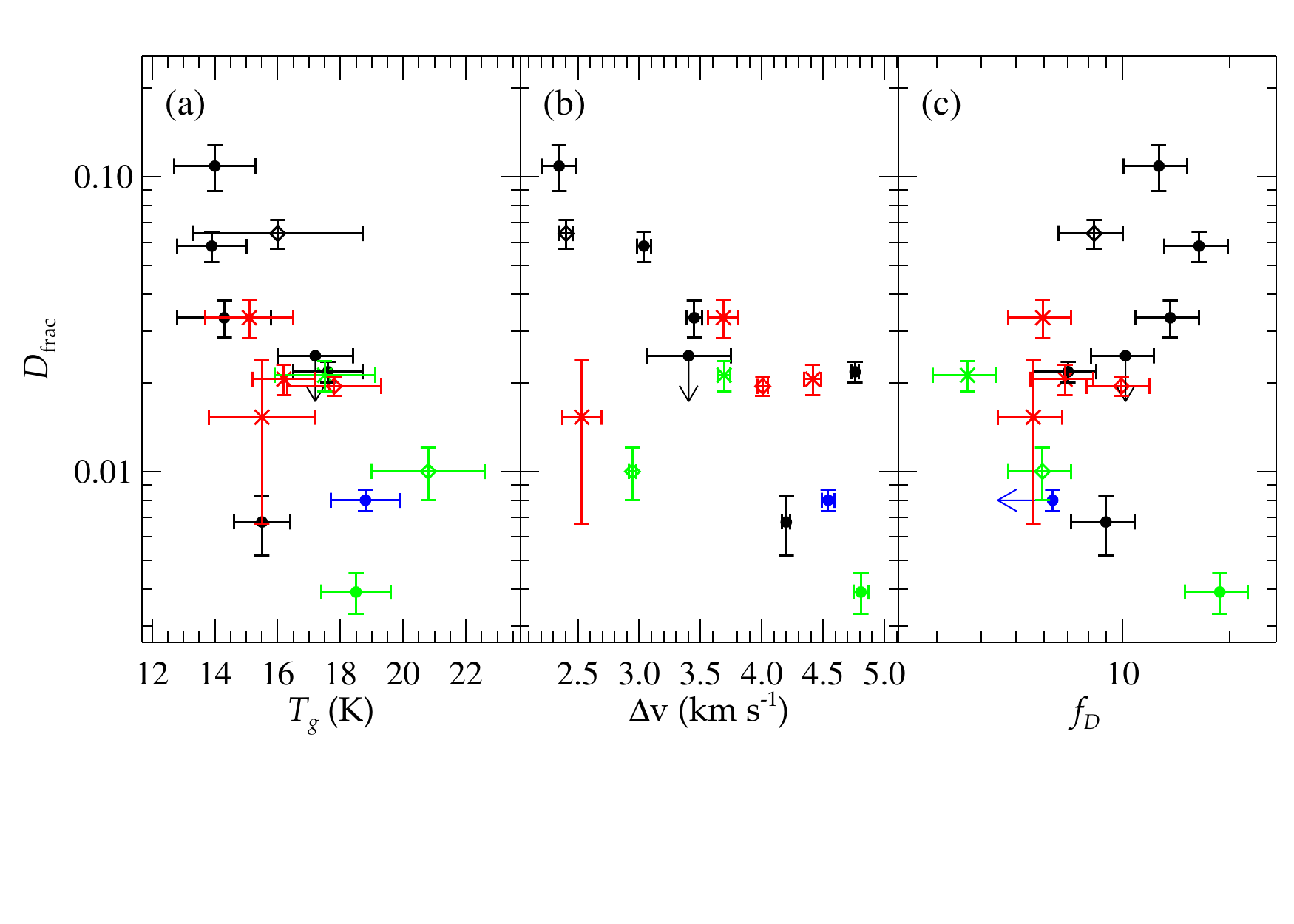} %
\caption{
(a) Deuterium fractionation, $D_\mathrm{frac}$, vs. gas temperature, $T_g$, shows a general decreasing trend.  
This correlation has Spearman's $\rho$ rank correlation coefficient of $\rho = -0.67$ with significance of $0.6 \, \%$ and Kendall's $\tau$ rank correlation coefficient of $\tau = -0.50$ with significance of $1.0 \, \%$.
Cores in G34.43, I18151, and I18223 are marked by filled circles, open diamonds, and crosses, respectively.  
Colors differentiate objects with known categories: blue for one UC \HII\ region, green for one HMC and HMPOs, and red for HMSCs; black color reserves for objects that were not previously classified. 
(b) $D_\mathrm{frac}$ vs. fitted linewidth, $\Delta v$, exhibits a decreasing trend that has $\rho = -0.61$ with significance of $1.6 \, \%$ and $\tau = -0.49$ with significance of $1.2 \, \%$. 
(c) $D_\mathrm{frac}$, vs. CO depletion factor, $f_D$.
Excluding G34.43-MM1 that has an unusually large $f_D$, the distribution of all other cores shows a weaker increasing trend that has $\rho = 0.49$ with significance of $7.8 \, \%$ and $\tau = 0.34$ with significance of $9.0 \, \%$.   
\label{f4}} %

\end{figure}

Except G34.43--MM8 with undetected $\mathrm{N_2D^+}$ emission, all our cores show a general   enhancement of 2$-$3 orders of magnitudes in $D_\mathrm{frac}$ (Table~\ref{t2}) over the local interstellar value of $1.51 \times 10^{-5}$ \markcite{Oliveira2003ApJ587}(Oliveira {et~al.} 2003).  
Note that the $D_\mathrm{frac}$ in G34.43 spans a fairly large range from $0.0039$ to $0.11$, nearly a factor of $30$.    
The deuterium fractionation, $D_\mathrm{frac}$, are compared with the gas temperature, $T_g$, the fitted linewidth, $\Delta v$, and the CO depletion factor, $f_D$ (Fig.~\ref{f4}). 
A clear decreasing trend in $D_\mathrm{frac}$ with both $T_g$ and $\Delta v$ but a weaker increasing trend with $f_D$ can be seen.  
Although these behaviors generally agree with expectations based on chemical models, an analytical formula to describe the dependence is not obvious with the large scatters in Fig.~\ref{f4}.  
To search for dependence, we perform statistical tests between $D_\mathrm{frac}$ and other parameters to evaluate the correlation along with the significance, which gives the likelihood for the correlation occurring by chance.  
The decreasing trend between $D_\mathrm{frac}$ and $T_g$ (Fig.~\ref{f4}a) has Spearman's $\rho$  rank correlation coefficient of $\rho = -0.67$ with significance of $0.6 \, \%$ and Kendall's $\tau$ rank correlation coefficient of $\tau = -0.50$ with significance of $1.0 \, \%$ while the correlation between $D_\mathrm{frac}$ and $\Delta v$ (Fig.~\ref{f4}b) gives $\rho = -0.61$ with significance of $1.6 \, \%$ and $\tau = -0.49$ with significance of $1.2 \, \%$.  
On the other hand, the dependence between $D_\mathrm{frac}$ and $f_D$ shows larger scatters (Fig.~\ref{f4}c).  
When excluding G34.43-MM1 with an unusually large $f_D$, we find an improved correlation that has $\rho=0.49$ with significance of $7.8 \, \%$ and $\tau = 0.34$ with significance of $9.0 \, \%$.
The concerns to include the HMC G34.43-MM1 will be elaborated in Sect.~\ref{subsec_fD}.  

\subsection{Deuterium Fractionation as an Evolutionary Probe \label{subsec_Df}} %

Over all, a monotonically decreasing trend in $D_\mathrm{frac}$ with both increasing gas temperature, $T_g$, and fitted linewidth, $\Delta v$, is discerned (Fig.~\ref{f4}a and \ref{f4}b).  
While examining cores in individual clouds, there seems to have a slightly better correlation, implying  possible cloud-to-cloud variation due to the influence of environments as previously suggested in the studies of low-mass cores \markcite{Crapsi2005ApJ619,Emprechtinger2009A&A493}(Crapsi {et~al.} 2005; Emprechtinger {et~al.} 2009).   
\markcite{Miettinen:2011um}Miettinen {et~al.} (2011) also suggest relatively large environmental variations in the cosmic-ray  ionization rates in massive IRDC cores. 
In particular, quiescent cores with no outflow activities, i.e. G34.43-MM6, -MM9, I18151-MM3, and I18223-MM2, all have the lowest temperature (Table~\ref{t1}) and the largest $D_\mathrm{frac}$ (Table~\ref{t2}) as well.  
Since warm and turbulent gas is more likely to be associated with evolved cores, the observed $D_\mathrm{frac}$ suggests a decreasing dependence with evolutionary stage.  
For a better determination of their evolutionary stage, one may desire to compare with theoretical evolutionary models, which depend on several physical parameters, such as bolometric temperature, bolometric luminosity, and envelope mass \markcite{Froebrich2005ApJS156}(Froebrich 2005).  
However, most of our selected cores, especially those with the largest $D_\mathrm{frac}$, do not have observations in the mid- and far-IR wavebands to better constrain their dust temperature and bolometric luminosity.  
Alternatively, we compare $D_\mathrm{frac}$ with gas temperature and linewidth instead of bolometric temperature and luminosity.  

A few previous studies have reported an anti-correlation between deuterium fractionation and evolutionary stage of massive proto-stellar cores.
\markcite{Chen2010ApJ713}Chen {et~al.} (2010) found $D_\mathrm{frac} = 0.017 - 0.052$ in three cores at different evolutionary stages within the IRDC G28.34+0.06, including a massive starless core (MM9), a core with fragmentation and outflow activities \markcite{Wang:2011jl}(MM4; Wang {et~al.} 2011), and an UC \HII\ region (MM1), and suggested a decreasing trend in $D_\mathrm{frac}$ with evolutionary stage.  
A subsequent study by \markcite{Fontani2011A&A529}Fontani {et~al.} (2011) significantly improved the statistics with a sample of twenty-seven cores and also revealed this decreasing trend with values of $D_\mathrm{frac}$ in the range of $0.012-0.7$, $0.017 - \le 0.4$, and $0.017-0.08$ for their observed HMSCs, HMPOs, and UC \HII\ regions, respectively.  
In particular, they found an anticorrelation between $D_\mathrm{frac}$ and $T_g$. 
\markcite{Miettinen:2011um}Miettinen {et~al.} (2011) further reported a decreasing trend in the ratio of $N(\mathrm{DCO^+})/N(\mathrm{HCO^+}) = 0.0002 - 0.014$ with gas temperature in their sample of seven IRDC cores. 
They also reported $D_\mathrm{frac} = 0.002 - 0.028$ for the four cores with higher gas temperature.    
Our values of $D_\mathrm{frac}$ are in the range of $0.0039 - 0.11$, which are comparable to the values obtained by \markcite{Fontani2011A&A529}Fontani {et~al.} (2011) and \markcite{Miettinen:2011um}Miettinen {et~al.} (2011).  
The anti-correlation between $D_\mathrm{frac}$ and $T_g$ is also seen in our results (Fig.~\ref{f4}a).  
 
\begin{figure} %
\plotone{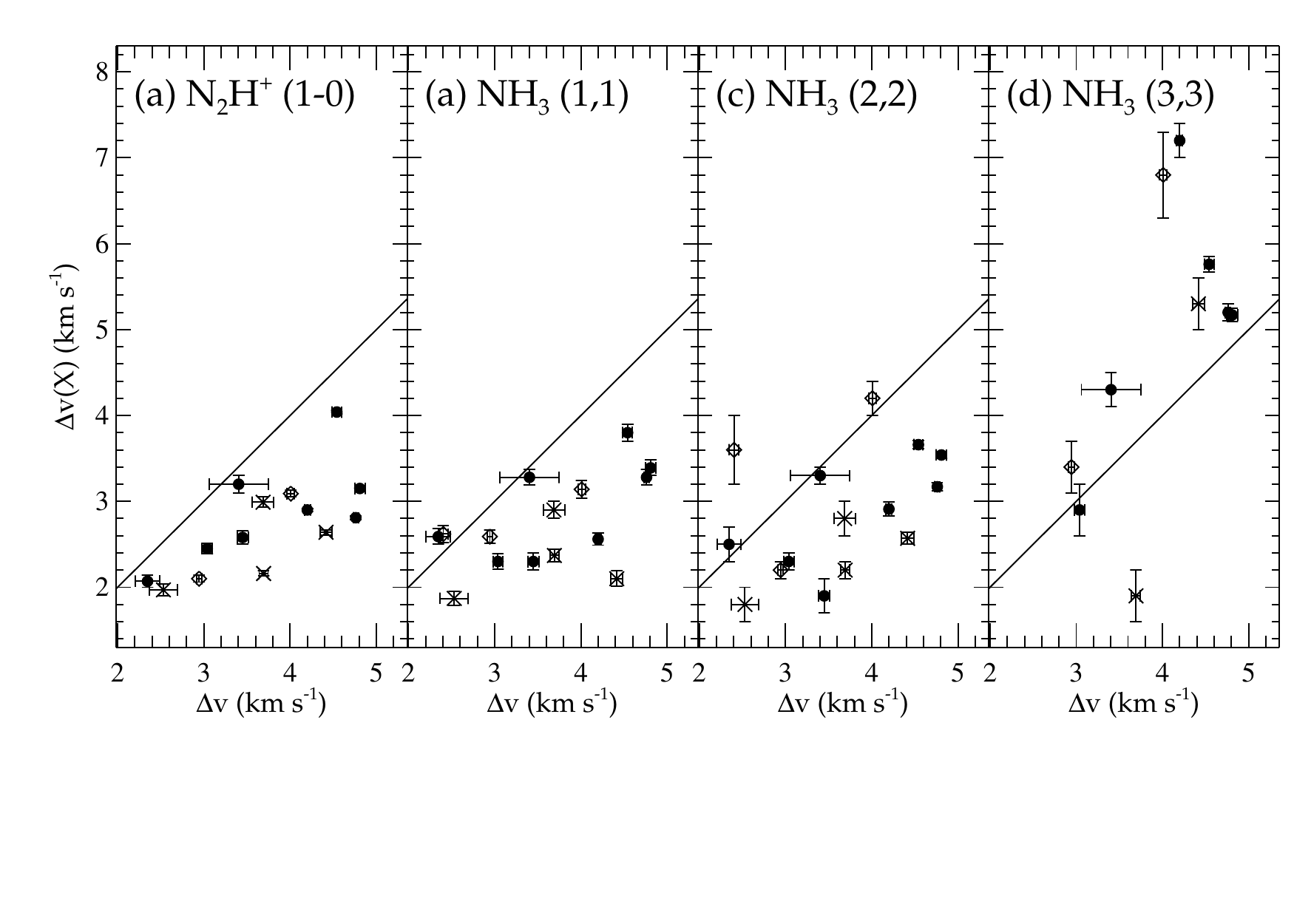} %
\caption{
(a) Linewidth $\mathrm{N_2H^+} \; (1-0)$ spectra from \markcite{Sakai2008ApJ678}Sakai {et~al.} (2008), $\Delta v(X)$, vs. linewidth of our $\mathrm{N_2H^+} \; (3-2)$ spectra, $\Delta v$.   
The angular resolutions for the $J = 1-0$ and $3-2$ observations are 18\arcsec\ and 27\arcsec, respectively.  
All the cores have larger linewidth in the $J=3-2$ transition.  
Cores in G34.43, I18151, and I18223 are marked by filled circles, open diamonds, and crosses, respectively. 
A solid line gives the locus of equal linewidths between the two spectra.  
The ordinate label, $\Delta v(X)$, stands for the linewidth of the line shown on the top.  
(b-d) Linewidth of $\mathrm{NH_3} \; (1,1)$, $(2,2)$, and $(3,3)$ spectra from \markcite{Sakai2008ApJ678}Sakai {et~al.} (2008) vs. linewidth of our $\mathrm{N_2H^+} \; (3-2)$ spectra.   
The angular resolution for $\mathrm{NH_3}$ observations are roughly 73\arcsec.
Except in the most quiescent cores, G34.43-MM6 and I18151-MM3, linewidths of the $\mathrm{N_2H^+} \; (3-2)$ spectra are larger than those of $\mathrm{NH_3} \; (1,1)$ and $(2,2)$ spectra.   
On the other hand, the $\mathrm{NH_3} \; (3,3)$ transition with higher $E_\mathrm{up} = 124.5 \; \mathrm{K}$ seems to tracer warmer component with linewidth larger than $\mathrm{N_2H^+} \; (3-2)$. 
Note that only ten cores, not including any of the four quiescent ones, were detected in $\mathrm{NH_3} \; (3,3)$ emission \markcite{Sakai2008ApJ678}(Sakai {et~al.} 2008).
\label{f5}} %
\end{figure}

Given the gas temperature range of $T_g = 14 - 21 \; \mathrm{K}$, the corresponding thermal linewidth of $\mathrm{N_2H^+}$ is merely $\Delta v_\mathrm{th} = 0.15 - 0.18 \; \mathrm{km \, s^{-1}}$. 
The observed linewidth in the range of $2.3 - 4.8 \; \mathrm{km \, s^{-1}}$ is dominated by non-thermal motions, possibly arising from turbulent motions among clumps.  
In Fig.~\ref{f5}, we compare the linewidth of our $\mathrm{N_2H^+} \; (3-2)$ spectra with linewidths of $\mathrm{N_2H^+} \; (1-0)$ and $\mathrm{NH_3} \; (1,1)$, $(2,2)$, and $(3,3)$ spectra observed by \markcite{Sakai2008ApJ678}Sakai {et~al.} (2008).  
The $\mathrm{N_2H^+} \; (1-0)$ observations had a smaller beam of 18\arcsec\ while the $\mathrm{NH_3}$ observations had a much larger beam of 73\arcsec\ with respect to our 27\arcsec\ beam for  $\mathrm{N_2H^+} \; (3-2)$.   
Between the two transitions of $\mathrm{N_2H^+}$, the $J=3-2$ transition with higher $E_\mathrm{up} = 26.8 \; \mathrm{K}$ shows broader linewidth than the $J=1-0$ with lower $E_\mathrm{up} = 4.5 \; \mathrm{K}$ as expected for warmer gas to be more turbulent (Fig.~\ref{f5}a).   
In general, $\mathrm{NH_3}$ lines have a much lower critical density of $n_\mathrm{crit} \sim 2 \times 10^3 \; \mathrm{cm^{-3}}$ \markcite{Evans1999ARA&A37}(Evans 1999) and tend to trace the outer part of the cores with respect to $\mathrm{N_2H^+}$ lines with $n_\mathrm{crit} \sim 10^5 \; \mathrm{cm^{-3}}$ . 
Except in the two quiescent cores, G34.43-MM6 and I18151-MM3, linewidths of the $\mathrm{NH_3} \; (1,1)$ and $(2,2)$ spectra are comparable to those of $\mathrm{N_2H^+} \; (1-0)$ but smaller than those of $\mathrm{N_2H^+} \; (3-2)$ (Fig.~\ref{f5}b and \ref{f5}c). 
The two quiescent cores are probably in a very early stage where turbulence dissipation may occur to produce smaller linewidth in the inner part of higher density \markcite{Goodman1998ApJ504}(Goodman {et~al.} 1998).  
In a number of more evolved cores, the $\mathrm{NH_3} \; (3,3)$ emissions with much higher $E_\mathrm{up} = 124.5 \; \mathrm{K}$ are detected and show much larger linewidth, even up to $7.2 \; \mathrm{km \, s^{-1}}$ in the case of G34.43-MM3 (Fig.~\ref{f5}d).  
Since all these evolved cores are associated with outflow activities, the $\mathrm{NH_3} \; (3,3)$ emission is tracing the hot gas which could be in the outflows \markcite{Zhang2002ApJ566}(Zhang {et~al.} 2002).  


Unlike $\mathrm{NH_3}$ and $\mathrm{NH_2D}$, which are affected by grain surface reactions \markcite{Gurtler2002A&A390,Bottinelli2010ApJ718}(G{\"u}rtler {et~al.} 2002; Bottinelli {et~al.} 2010), $\mathrm{N_2D^+}$ and $\mathrm{N_2H^+}$ are pure gas-phase reactants and do not participate condensation and subsequent sublimation of ice mantles.  
Compared to other molecular species, the deuterium fractionation of $\mathrm{N_2H^+}$ better reflects the physical conditions at present time without being confused by evaporation of mantles which had formed at earlier times with enhanced deuteration \markcite{Emprechtinger2009A&A493}(Emprechtinger {et~al.} 2009).  
In a sample of Taurus cores, a clear increasing trend in the deuterium fractionations of both $\mathrm{NH_3}$ and $\mathrm{N_2H^+}$ was observed in prestellar cores \markcite{Crapsi2005ApJ619,Hatchell2003A&A403}(Crapsi {et~al.} 2005; Hatchell 2003), whereas in protostellar cores, $D_\mathrm{frac}$  shows a faster decreasing trend with dust temperature than does the $N(\mathrm{NH_2D})/N(\mathrm{NH_3})$ ratio \markcite{Emprechtinger2009A&A493}(Emprechtinger {et~al.} 2009).  
For high-mass protostellar cores within similar gas temperature range, we note that the correlation between $D_\mathrm{frac}$ and $T_g$ appears stronger in our cores than does the $N(\mathrm{NH_2D})/N(\mathrm{NH_3})$ ratio \markcite{Pillai2007A&A467}(Pillai {et~al.} 2007).  

\subsection{Deuterium Fractionation and the CO Depletion Factor \label{subsec_fD}} %
To further examine the relationship between deuterium fractionation and the CO depletion, we computed the $\mathrm{C^{18}O}$ column density and the CO depletion factor, $f_D$, as described in Sect.~\ref{sec_fD}.
The results are listed in Table~\ref{t3} and shown in Fig.~\ref{f4}c.  
When excluding G34.43-MM1, which has the largest value of $f_D$, we find an increasing trend in $D_\mathrm{frac}$ with $f_D$.    
This agrees with the general expectations from chemical models that CO is the major destroyer of $\mathrm{H_3^+}$, $\mathrm{H_2D^+}$, $\mathrm{N_2H^+}$, and $\mathrm{N_2D^+}$ \markcite{Caselli:1998dv}(Caselli {et~al.} 1998).  
As the envelope heats up, the CO abundance is expected to quickly rise up based on the dramatical drop of the CO sublimation timescale from $10^8 \; \mathrm{yr}$ at $T_d \simeq 12 \; \mathrm{K}$ to $0.1 \; \mathrm{yr}$ at $T_d \simeq 20 \; \mathrm{K}$ \markcite{Collings2003Ap&SS285}(Collings {et~al.} 2003).  
The warmer temperature together with the return of gas-phase CO can alter the competition in the chemical networks and brings a drop in $D_\mathrm{frac}$ \markcite{Roueff2005A&A438,Aikawa2005ApJ620,Aikawa2008Ap&SS313}(Roueff {et~al.} 2005; Aikawa {et~al.} 2005; Aikawa 2008).  
When CO returns to the gas phase, it will quickly react with $\mathrm{N_2H^+}$ and $\mathrm{N_2D^+}$ to form $\mathrm{HCO^+}$ and $\mathrm{DCO^+}$, respectively \markcite{Lee2004ApJ617}(Lee, Bergin, \& Evans 2004). 
However, one should be cautious about possible chemical stratification once the CO sublimation starts  in the central warm region.  
As a core warms up, the $\mathrm{N_2H^+}$ and $\mathrm{N_2D^+}$ emissions tend to trace the cold outer region whereas the CO emission is dominated in the central region.  
In the study of the Ophiuchus B2 core,  \markcite{Friesen2010ApJ718}Friesen {et~al.} (2010) found that $D_\mathrm{frac}$ increases at greater projected distances from the embedded sources.  
When the observed emission arises from a partially filled volume, the beam-averaged abundance for each species will start to deviate from the actual abundance used in chemical models.  
Observations with high angular resolutions will help to image the spatial distributions and reduce the confusion in comparing abundances.  

In the case of G34.43-MM1, an unusually large value of $N_\mathrm{H_2}$ and hence $f_D$ is obtained.  
Millimeter interferometric studies \markcite{Cortes2008ApJ676,Rathborne2008ApJ689}(Cortes {et~al.} 2008; Rathborne {et~al.} 2008) reveal a very strong ($L_\mathrm{bol} \sim 2 \times 10^4 \, L_\odot$) and compact ($2 R \simeq 0.03 \; \mathrm{pc}$) source that exhibits signatures of an HMC, which is thought to have a typical temperature closer to $100 \; \mathrm{K}$.  
This source has developed a steep temperature gradient, from $34$ to $100 \; \mathrm{K}$ across core scales from $0.1$ to $0.015 \; \mathrm{pc}$ \markcite{Rathborne2008ApJ689}(Rathborne {et~al.} 2008), translating to a single power-law dependence of $r^{-0.57}$.   
This steep temperature gradient suggests the presence of an inner region where optical depth is large to the IR photons that the photon diffusion should be considered \markcite{Kenyon1993ApJ414a,Osorio1999ApJ525,Chen2006ApJ639}(Kenyon, Calvet, \& Hartmann 1993; Osorio, Lizano, \& D'Alessio 1999; Chen {et~al.} 2006).  
In our approach to estimate $f_D$, the dust emission is likely dominated by the hot inner region while the CO emission arises from a cold and large envelope.     
The dust temperature, $T_d$, derived from the SED fit depends on the flux densities in the mid-IR wavebands that may suffer from significant optical depth and does not reflect the physical conditions of the hot central region where the peak of the optically thin 1.2~mm emission is produced.  
Such steep temperature gradient may cause an underestimate in $T_d$ and overestimates in $N_\mathrm{H_2}$ and $f_D$ in our current calculation.  
To avoid potentially misleading interpretation, we exclude the result of G34.43-MM1 when discussing the electron abundance in the following section.  

\subsection{The Ionization Degree \label{subsec_xe}} %
The enrichment of primary deuterated ions, e.g. $\mathrm{H_2D^+}$, $\mathrm{CH_2D^+}$, and $\mathrm{C_2HD^+}$, will give rise to the enrichment of subsequent deuterated species, such as $\mathrm{N_2D^+}$, but with lower $\mathrm{[D]/[H]}$ abundance ratios due to the statistical nature of the fractionation process.  
In simple steady-state models based on gas-phase ion-molecular chemistry, an upper limit to the electron abundance, $x_e$, can be found with assumptions that all the deuterium enrichment originates in $\mathrm{H_2D^+}$ and that the recombination on negatively charged grains is negligible  \markcite{Wootten:1979il,Caselli:1998dv,Caselli:2002hf}(Wootten, Snell, \& Glassgold 1979; Caselli {et~al.} 1998; Caselli 2002).    
Following the method described in \markcite{Caselli:2002hf}Caselli (2002), the deuterium fractionation, $D_\mathrm{frac}$, may be expressed as a function of $x_e$ and abundances of $\mathrm{HD}$, $x(\mathrm{HD})$, and important neutral species, $x(m)$, 
\begin{equation} %
D_\mathrm{frac} 
= \frac{1}{3} \frac{k_\mathrm{HD} \, x(\mathrm{HD})}{k_e x_e + \displaystyle \sum_m k_m \, x(m), } 
\end{equation} %
where $k_\mathrm{HD} = 1.5 \times 10^{-9} \; \mathrm{cm^3 \, s^{-1}}$ is the rate coefficient for the reaction in Eq.~(\ref{eq_HD}), $k_e = 6 \times 10^{-8} \, (T/300)^{-0.65} \; \mathrm{cm^3 \, s^{-1}}$ the dissociated recombination rate of $\mathrm{H_2D^+}$ \markcite{Caselli:1998dv}(Caselli {et~al.} 1998), $k_m$ the destruction rate for $\mathrm{H_2D^+}$ due to reactions with neutral species $m$, such as CO and O.  
The numerical factor of $1/3$ accounts for the statistical branching ratio of $1/3$ to transfer the deuteron in the reaction of $\mathrm{H_2D^+}$ with $\mathrm{N_2}$.      
The $\mathrm{HD}$ abundance is taken from the interstellar value of $x(\mathrm{HD}) = 2 \mathrm{[D]/[H]} = 3 \times 10^{-5}$ \markcite{Oliveira2003ApJ587,Caselli:1998dv}(Oliveira {et~al.} 2003; Caselli {et~al.} 1998).  
The electron abundance $x_e$ is then given by 
\begin{equation} %
x_e = \frac{k_\mathrm{HD} \, x_\mathrm{HD}}{3 \, k_e} \frac{1}{D_\mathrm{frac}} - \frac{1}{k_e}  \sum_m k_m x(m).  
\label{eq_Dfxe}
\end{equation} %
Since CO is the dominant neutral species that destroys $\mathrm{H_2D^+}$, we make an approximation by neglecting other ion-neutral reactions to get 
\begin{equation} %
\sum_m k_m x(m) \gtrsim k_\mathrm{CO} \, x(\mathrm{CO}) = k_\mathrm{CO} \, \left( \frac{x(\mathrm{CO})_\mathrm{can}}{f_D} \right), 
\end{equation} %
where $k_\mathrm{CO} = 6 \times 10^{-10} \, (T/300)^{-0.5} \; \mathrm{cm^3 \, s^{-1}}$ is the $\mathrm{H_2D^+}$ destruction rate due to reactions with CO \markcite{Caselli:1998dv}(Caselli {et~al.} 1998), and $x(\mathrm{CO})_\mathrm{can} = 1.5 \times 10^{-4}$ is the canonical CO abundance at the locations of our cores.  
We may set an upper limit for $x_e$ to be 
\begin{eqnarray} %
x_e &\lesssim& \frac{k_\mathrm{HD} \, x_\mathrm{HD}}{3 \, k_e} \frac{1}{D_\mathrm{frac}} - \frac{k_\mathrm{CO} \, x(\mathrm{CO})_\mathrm{can}}{k_e} \frac{1}{f_D} \\ 
&=& \frac{3.8 \times 10^{-8}}{D_\mathrm{frac}} - \frac{9.7 \times 10^{-7}}{f_D} \hspace{3cm} (\mathrm{for} \;\; T_g = 16 \; \mathrm{K}).  
\end{eqnarray} %
For easy comprehension of the dependence on $D_\mathrm{frac}$ and $f_D$, the numerical values are provided for the case of $T_g = 16 \; \mathrm{K}$.  
With the derived $D_\mathrm{frac}$ and $f_D$, we obtain the ionization degree in the range of $x_e = 2 \times 10^{-7} - 5 \times 10^{-6}$ for our selected cores (Table~\ref{t3}).  
These values lie at high end of the ionization degrees reported in early studies of low-mass dense cores \markcite{Caselli:1998dv,Williams:1998fa}(Caselli {et~al.} 1998; Williams {et~al.} 1998) and massive cores \markcite{Bergin:1999cw}(Bergin {et~al.} 1999).   
Recently, \markcite{Miettinen:2011um}Miettinen {et~al.} (2011) derived first estimates for ionization degrees in IRDC cores with  upper limits of $x_e = 2 \times 10^{-6} - 2.9 \times 10^{-4}$ and lower limits of $x_e = 3 \times 10^{-9} - 5.6 \times 10^{-8}$.  
Our estimates give smaller values compared to their upper limits but are within the range bracketed by their upper and lower limits.  
The smaller $x_e$ upper limits are mainly attributed to the larger deuterium fractionation derived from $\mathrm{N_2H^+}$ instead of $\mathrm{HCO^+}$.  
Furthermore, our $x_e$ values show a moderate correlation with the evolutionary stage.  
Since more evolved cores show smaller values of $D_\mathrm{frac}$ and $f_D$, the plausible  correlation between $x_e$ and evolutionary stage may lead to the decreasing trend of $D_\mathrm{frac}$.  
Most of the cores in our sample have shown outflow signatures and likely have begun to form clusters with low- and intermediate-mass protostars.  
This increasing degree of ionization could be arising from accretion shocks and/or heating of the gas related to the central proto-stellar/cluster objects \markcite{Stahler1980ApJ242,Hosokawa2010ApJ721,Calvet2004AJ128}(Stahler, Shu, \& Taam 1980; Hosokawa, Yorke, \&  Omukai 2010; Calvet {et~al.} 2004).  
Given the porous nature of the surrounding medium \markcite{Indebetouw2006ApJ636}(Indebetouw {et~al.} 2006), it is possible for part of the energetic photons to reach outer part of the core and rise up the overall electron abundance \markcite{Kim2001ApJ549}(Kim \& Koo 2001).  

The four quiescent cores, G34.43-MM6, -MM9, I18151-MM3, I18223-MM2, also have the lowest ionization degree, $x_e = 2 \times 10^{-7} - 9 \times 10^{-7}$, among cores in same IRDC.  
In these cores, the degree of ionization may play a role in regulating star formation efficiency through ambipolar diffusion, in which magnetic fields drift relative to a background of neutrals \markcite{Mestel:1956we,Shu1987ARA&A25}(Mestel \& Spitzer 1956; Shu, Adams, \& Lizano 1987).  
In a partially ionized medium, charged particles are coupled to magnetic fields while neutrals are supported against their self-gravity through the frictional drag that they experience when drifting through ions.   
In addition to ambipolar diffusion, evolution of massive star-forming cores can be affected by various effects such as turbulence \markcite{McKee2003ApJ585}(McKee \& Tan 2003), rotation, and magnetic fields.
For a more complete picture, one needs to consider all the important supports against gravity \markcite{Myers:1988hb,McKee2007ARA&A}(Myers \& Goodman 1988; McKee \& Ostriker 2007).    
Here we just make an attempt to estimate the characteristic scale for ambipolar diffusion, $\ell_\mathrm{AD}$, which gives the smallest scale for which the magnetic field is well coupled to the bulk of the gas.
Following the picture described in \markcite{McKee2007ARA&A}McKee \& Ostriker (2007) and approximating the ion abundance with $n_i = x_e \, n_\mathrm{H_2}$, we estimate the characteristic ambipolar diffusion scale to be  
\begin{eqnarray} %
\ell_\mathrm{AD} = \frac{v_A}{n_i \, \alpha_{in}} &\approx & 0.05 \; \mathrm{pc} \left( \frac{v_A}{3 \; \mathrm{km \, s^{-1}}} \right) \left( \frac{n_i}{10^{-3} \; \mathrm{cm^{-3}}} \right)^{-1} \nonumber \\
&=& 0.002 \; \mathrm{pc} \left( \frac{B}{10 \; \mathrm{\mu G}} \right) \left( \frac{x_e}{10^{-7}} \right)^{-1} \left( \frac{n_\mathrm{H_2}}{10^4 \; \mathrm{cm^{-3}}} \right)^{-3/2}, 
\end{eqnarray} %
where $v_A = B/\sqrt{4 \pi \, \mu \, m_\mathrm{H_2} \, n_\mathrm{H_2} }$ is the Alfv\'{e}n speed associated with the large-scale magnetic field $B$ and $\alpha_{in} \approx 2 \times 10^{-9} \; \mathrm{cm^3 \, s^{-1}}$ the ion-neutral collision rate coefficient \markcite{Draine:1983jz}(Draine, Roberge, \& Dalgarno 1983).  
For the unknown magnetic field strength in supersonic cores, we take the median value given by \markcite{McKee2007ARA&A}McKee \& Ostriker (2007) (originally from \markcite{Crutcher1999ApJ520}Crutcher (1999)),   
\begin{equation} %
B_\mathrm{med} = 30 \left( \frac{2 \, n_\mathrm{H_2}}{10^3 \; \mathrm{cm^{-3}}} \right)^{1/2} \left( \frac{\Delta v_\mathrm{nt}/\sqrt{8 \ln 2}}{1 \; \mathrm{km \, s^{-1}}} \right) \; \mathrm{\mu G},    
\end{equation} %
where $\Delta v = \sqrt{\Delta v^2 - \Delta v_\mathrm{th}^2}$ is the line width for nonthermal motions.  
Applying the values for $n_\mathrm{H_2}$ and $\Delta v$ in the four quiescent cores, we find the magnetic field strength to be in a range from $210$ to $572 \; \mathrm{\mu G}$.  
The characteristic scale for ambipolar diffusion is in a range from $\ell_\mathrm{AD} = 6 \times 10^{-4}$ to $4 \times 10^{-3} \; \mathrm{pc}$, much smaller than the typical size of the cores.

Since most star-forming cores have  Alfv\'{e}n speeds comparable to the nonthermal component of the velocity dispersions, the turbulence in these cores should have a substantial magnetohydro-dynamic (MHD) component \markcite{Myers:1988hb,Crutcher1999ApJ514}(Myers \& Goodman 1988; Crutcher {et~al.} 1999).
\markcite{Myers1998ApJ507}Myers \& Lazarian (1998) have proposed a pressure-driven cooling flows associated with local dissipation of turbulence due to wave damping by ion-neutral friction in an inner core region.    
From a sufficiently turbulent outer region, the MHD wave power transmission, $g$, into a spherical inner core is described in terms of the field-neutral coupling parameter, $W$, which is defined as the ratio of the core size, $R$, to the minimum cutoff wavelength, $\lambda_0 = \pi \ell_\mathrm{AD}$, for the propagation of MHD waves \markcite{Myers:1995ff}(Myers \& Khersonsky 1995). 
Using the upper limits of $x_e$ in the four quiescent cores, we obtain $W = 23 - 119$ with a mean value of $W = 75$, roughly in the regime of marginal field-neutral coupling \markcite{Myers1998ApJ507}($g \sim 0.4$ for $W \lesssim 100$; Myers \& Lazarian 1998).  
The MHD waves excited in the outer region would have a fairly limited range of allowed wavelengths to transmit the turbulence power into the dissipating inner regions, and a pressure-driven turbulent cooling flow may occur in these cores.   
For comparison, \markcite{Bergin:1999cw}Bergin {et~al.} (1999) reported $W=20$ in massive cores with ionization degree inferred from chemical models while \markcite{Miettinen:2011um}Miettinen {et~al.} (2011) obtained $W \le 18.5$ using their lower limits of ionization degree.  
But one should be cautious about a direct comparison among these values because of different methods to estimate the ionization degrees and the magnetic field strength.  

\section{Summary} %
We observed emissions of $\mathrm{N_2H^+} \; (3-2)$, $\mathrm{N_2D^+} \; (3-2)$, and $\mathrm{C^{18}O \; (2-1)}$ toward 15 cores in the IRDC G34.43, and the HMPO I18151 and I18223.
The main findings are summarized as follows:
\begin{enumerate} %

\item A clear decreasing trend in the deuterium fractionation of $\mathrm{N_2H^+}$, $D_\mathrm{frac}$, with evolutionary stage traced by increasing gas temperature and linewidth.   
This decreasing trend agrees with the findings in previous studies by \markcite{Chen2010ApJ713}Chen {et~al.} (2010), \markcite{Fontani2011A&A529}Fontani {et~al.} (2011), and \markcite{Miettinen:2011um}Miettinen {et~al.} (2011).    
An increasing trend, though with larger scatters, in $D_\mathrm{frac}$ with the CO depletion factor, $f_D$, is also found.  
Such trend resembles the behavior of $D_\mathrm{frac}$ in the low-mass protostellar cores and suggests the use of the $N(\mathrm{N_2D^+})/N(\mathrm{N_2H^+})$ ratio as an evolutionary probe to high-mass proto-stellar/cluster candidates.  

\item A significant enhancement of $2-3$ orders of magnitude in the $N(\mathrm{N_2D^+})/N(\mathrm{N_2H^+})$ ratio in all detected sources over the local interstellar $\mathrm{[D]/[H]}$ ratio.   Such enhancement agree well with those observed in other massive star-forming cores.   

\item The upper limits of electron abundance are estimated to be in the range from $2 \times 10^{-7}$ to $5 \times 10^{-6}$, which lie at the high end of the typical values observed in early studies but within the range found by \markcite{Miettinen:2011um}Miettinen {et~al.} (2011) in their IRDC cores.  
More evolved cores seem to show higher degree of ionization, which may be related to star-forming activities.  

\item In the four quiescent cores, the inferred characteristic scale for ambipolar diffusion is roughly $10^{-3} \; \mathrm{pc}$, and the coupling parameter of $W \lesssim 120$ are within the regime of marginal field-neutral coupling.  
The physical conditions may favor turbulent cooling flows.  
\end{enumerate} %
 
\acknowledgments %
We thank Dr. H.~Beuther and Dr. J.~M. Rathborne for providing the 1.2~mm MAMBO and MAMBO2 maps.  
We also thank Dr. H.~Shang and Dr. R.~Krasnopolsky for the helpful discussion on ambipolar diffusion.  
This research is supported by National Science Council of Taiwan through grants NSC 97-2112-M-007-006-MY3 and NSC 100-2112-M-007-004-MY2.  


\appendix %
\section{Discussions on Individual Clouds} %
G34.43+0.24.--- The IRDC G34.43 ($d = 3.7 \; \mathrm{kpc}$) contains nine cores \markcite{Rathborne2006ApJ641}(Rathborne {et~al.} 2006) with G34.43-MM2 being the most evolved core associated with the UC \HII\ region IRAS~18507+0121 of spectral type B0.5 \markcite{Molinari:1998wl,Shepherd2004ApJ602,Shepherd2007ApJ669}(Molinari {et~al.} 1998; Shepherd, N{\"u}rnberger, \&  Bronfman 2004; Shepherd {et~al.} 2007).  
The brightest millimeter core, G34.43-MM1, exhibits typical chemical signature of a hot molecular core (HMC) and has started internal heating with embedded source(s) \markcite{Rathborne2008ApJ689}(Rathborne {et~al.} 2008).  
Additionally, G34.43-MM1, MM3, MM4, MM5, and MM8 are associated with extended {\it Spitzer} $4.5 \; \mu\mathrm{m}$ emissions, indicating possibly outflow activities \markcite{Chambers2009ApJS181}(Chambers {et~al.} 2009).  

IRAS 18151$-$1208.--- The HMPO I18151 ($d = 3.0 \; \mathrm{kpc}$) hosts four dusty cores \markcite{Beuther2002ApJ566}(Beuther {et~al.} 2002a) with MM1 being the most evolved and dominant $K$-band source, possibly driving $\mathrm{H_2}$ jets and CO outflows \markcite{Beuther2002A&A383,Davis2004A&A425}(Beuther {et~al.} 2002b; Davis {et~al.} 2004).    
By analyzing molecular line emissions and the spectral energy distributions (SEDs), \markcite{Marseille2008A&A488}Marseille {et~al.} (2008) suggested an evolutionary sequence among three most compact cores: from the youngest I18151-MM3 perhaps in a pre-stellar phase, followed by MM2 as an HMSC with an embedded, mid-IR-quiet young protostar, to the most evolved MM1 with mid-IR-bright protostars.  
Molecular outflows in I18151-MM1 and MM2 also suggest their more evolved stages \markcite{Marseille2008A&A488,Beuther2007ApJ668}(Marseille {et~al.} 2008; Beuther \& Sridharan 2007).    

IRAS 18223$-$1243.--- The HMPO I18223 ($d = 3.7 \; \mathrm{kpc}$) harbors a few dusty cores in a filamentary structure with I18223-MM1 most evolved as a HMPO and others as HMSCs \markcite{Sridharan2005ApJ634}(Sridharan {et~al.} 2005).   
Large SiO linewidths suggest outflow activities in I18223-MM3 and MM4 \markcite{Beuther2007ApJ668}(Beuther \& Sridharan 2007). 



\begin{deluxetable}{ccccccccc} %
\tablewidth{0pt} %
\tablecolumns{9} %
\tablecaption{Properties of the Observed Cores \label{t1}} %
\tablehead{ \colhead{Core} & \colhead{R.A.} & \colhead{Decl.} &  \colhead{$T_g$\tablenotemark{a}} & \colhead{$F_\mathrm{1.2~mm}$\tablenotemark{b}} & \colhead{$2 R$\tablenotemark{b}} & \colhead{$\log n_\mathrm{H_2}$} & \colhead{Outflow\tablenotemark{c}} &\colhead{Remark\tablenotemark{d}}  \\
\colhead{} & \colhead{(J2000)} & \colhead{(J2000)} & \colhead{(K)} & \colhead{(Jy)} & \colhead{(pc)} & \colhead{($\mathrm{cm^{-3}}$)} & \colhead{(Y/N)} & } %
\startdata %
\sidehead{G34.43+0.24 (IRDC; $d = 3.7 \; \mathrm{kpc}$)} \cline{1-3} %
MM1 & 18 53 18.0 & +01 25 24 & 18.5 & 4.01 & 0.19 & 6.44 & Y & HMC \\
MM2 & 18 53 18.6 & +01 24 40 & 18.8 & 4.33 & 0.42 & 5.42 & Y & UC \HII \\
MM3 & 18 53 20.4 & +01 28 23 & 15.5 & 1.02 & 0.38 & 4.95 & Y & \\
MM4 & 18 53 19.0 & +01 24 08 & 17.6 & 0.86 & 0.38 & 4.87 & Y & \\
MM5 & 18 53 19.8 & +01 23 30 & 14.3 & 2.24 & 0.89 & 4.65 & Y & \\
MM6 & 18 53 18.6 & +01 27 48 & 14.0 & 0.43 & 0.62 & 4.41 & N & \\
MM8 & 18 53 16.4 & +01 26 20 & 17.2 & 0.36 & 0.52 & 4.44 & Y & \\
MM9 & 18 53 18.4 & +01 28 14 & 13.9 & 0.53 & 0.67 & 4.41 & N & \\
\hline
\sidehead{IRAS~18151-1208 (HMPO; $d = 3.0 \; \mathrm{kpc}$)} \cline{1-3} %
MM1 & 18 17 58.0 & $-12$ 07 27 & 20.8 & 3.6 & 0.25 & 5.94 & Y & HMPO \\
MM2 & 18 17 50.4 & $-12$ 07 55 & 17.8 & 2.6 & 0.34 & 5.62 & Y & HMSC \\
MM3 & 18 17 52.2 & $-12$ 06 56 & 16.0 & 0.9 & 0.40 & 5.05 & N & \\
\hline
\sidehead{IRAS~18223-1243 (HMPO; $d = 3.7 \; \mathrm{kpc}$)} \cline{1-3} %
MM1 & 18 25 10.5 & $-$12 42 26 & 17.5 & 2.5 & 0.55 & 4.88 & Y & HMPO \\
MM2 & 18 25 09.5 & $-$12 44 15 & 15.1 & 0.6 & 0.47 & 4.88 & N & HMSC \\
MM3 & 18 25 08.3 & $-$12 45 28 & 16.2 & 0.8 & 0.31 & 5.42 & Y & HMSC \\
MM4 & 18 25 07.2 & $-$12 47 54 & 15.5 & 0.3 & 0.52 & 4.43 & Y & HMSC
\enddata %
\tablenotetext{a}{~Gas temperature based on $\mathrm{NH_3}$ observations \markcite{Sakai2008ApJ678}(Sakai {et~al.} 2008).}
\tablenotetext{b}{Integrated 1.2~mm flux density and the deconvolved core size, defined as the geometric mean of the major and minor FWHMs. \markcite{Beuther2002ApJ566,Rathborne2006ApJ641}(Beuther {et~al.} 2002a; Rathborne {et~al.} 2006). } %
\tablenotetext{c}{Compiled from literature using {\it Spitzer} $4.5 \; \mu\mathrm{m}$ emission and molecular outflows \markcite{Beuther2002A&A383,Chambers2009ApJS181,LopezSepulcre2011A&A526,Marseille2008A&A488}(Beuther {et~al.} 2002b; Chambers {et~al.} 2009; L{\'o}pez-Sepulcre {et~al.} 2011; Marseille {et~al.} 2008).} %
\tablenotetext{d}{Classification from previous studies \markcite{Rathborne2006ApJ641,Rathborne2008ApJ689,Sridharan2002ApJ566,Sridharan2005ApJ634}(Rathborne {et~al.} 2006, 2008; Sridharan {et~al.} 2002, 2005). } %
\end{deluxetable} %

\begin{deluxetable}{ccccccccc} %
\tablewidth{0pt} %
\tablecolumns{9} %
\tablecaption{Parameters of Spectral Fits for $\mathrm{N_2H^+}$ and $\mathrm{N_2D^+}$ \label{t2}} %
\tablehead{ \colhead{Core} & \colhead{$\sigma_\mathrm{N_2H^+}$} & \colhead{$\sigma_\mathrm{N_2D^+}$} & \colhead{$N(\mathrm{N_2H^+})$} & \colhead{$\upsilon_\mathrm{LSR}$} & \colhead{$\Delta \upsilon$} & \colhead{$D_\mathrm{frac}$} & \colhead{$\overline{\chi^2}$} & \colhead{$\tau_\mathrm{max}$} \\
\colhead{} & \colhead{(mK)} & \colhead{(mK)} & \colhead{($10^{12} \; \mathrm{cm^{-2}}$)} & \colhead{($\mathrm{km \, s^{-1}}$)} & \colhead{($\mathrm{km \, s^{-1}}$)} & \colhead{} & \colhead{} & \colhead{} } %
\startdata %
\sidehead{G34.43+0.24} \cline{1-3}  
MM1 & 44 & 9.5 & $17\pm2$ & $57.550\pm0.008$ & $4.81\pm0.06$ & $(39\pm6) \times 10^{-4}$ & 3.7 & 0.77  \\
MM2 & 20 & 7.7 & $13\pm1$ & $57.579\pm0.005$ & $4.54\pm 0.05$ & $(80\pm7) \times 10^{-4}$ & 9.8 & 0.61 \\
MM3 & 20 & 7.8 & $6.0\pm0.6$ & $59.66\pm0.01$ & $4.20\pm0.03$ & $0.007\pm0.002$ & 3.1 & 0.32 \\
MM4 & 21 & 8.8 & $5.8\pm0.5$ & $57.56\pm0.01$ & $4.76\pm0.03$ & $0.022\pm0.002$ & 4.2 & 0.27 \\
MM5 & 19 & 7.1 & $1.8\pm0.3$ & $57.89\pm0.03$ & $3.45\pm0.06$ & $0.033\pm0.005$ & 1.6 & 0.12 \\
MM6 & 23 & 10.5 & $0.59\pm0.09$ & $58.27\pm0.06$ & $2.3\pm0.1$ & $0.11\pm0.02$ & 0.8 & 0.05 \\
MM8 & 24 & 8.5 & $0.33\pm0.04$ & $57.2\pm0.1$ & $3.4\pm0.3$ & $0.00\pm0.02$ & 1.2 & 0.02 \\
MM9 & 17 & 8.9 & $1.6\pm0.2$ & $58.89\pm0.02$ & $3.04\pm0.06$ & $0.058\pm0.007$ & 2.5 & 0.11 \\
\hline
\sidehead{IRAS~18151-1208} \cline{1-3}  
MM1 & 25 & 10.6 & $3.9\pm0.4$ & $33.203\pm0.008$ & $2.94\pm0.03$ & $0.010\pm0.002$ & 6.0 & 0.27 \\
MM2 & 22 & 7.9 & $6.6\pm0.8$ & $29.706\pm0.007$ & $4.01\pm0.04$ & $0.019\pm0.001$ & 3.9 & 0.36 \\
MM3 & 20 & 8.3 & $1.1\pm0.3$ & $30.67\pm0.02$ & $2.40\pm0.06$ & $0.064\pm0.007$ & 2.4 & 0.10 \\
\hline 
\sidehead{IRAS~18223-1243} \cline{1-3}
MM1 & 30 & 8.6 & $3.5\pm0.4$ & $45.41\pm0.02$ & $3.69\pm0.05$ & $0.021\pm0.002$ & 3.4 & 0.20 \\
MM2 & 38 & 7.9 & $1.9\pm0.3$ & $45.20\pm0.05$ & $3.7\pm0.1$ & $0.033\pm0.005$ & 1.1 & 0.11 \\
MM3 & 40 & 8.6 & $4.2\pm0.4$ & $45.54\pm0.03$ & $4.42\pm0.07$ & $0.021\pm0.002$ & 1.5 & 0.21 \\
MM4 & 43 & 8.4 & $0.9\pm0.2$ & $45.93\pm0.07$ & $2.5\pm0.2$ & $0.015\pm0.009$ & 1.2 & 0.08 \\
\enddata %
\end{deluxetable} %

\begin{deluxetable}{cccccccc} %
\tablewidth{0pt} %
\tablecolumns{8} %
\tablecaption{Parameters of C$^{18}$O Spectra and Derived Quantities \label{t3}} %
\tablehead{ \colhead{Core} & \colhead{$\sigma_\mathrm{C^{18}O}$} & \colhead{$W(\mathrm{C^{18}O})$} & \colhead{$N(\mathrm{C^{18}O})$} & \colhead{$T_d$\tablenotemark{a}} & \colhead{$S_\mathrm{1.2 \, mm}$} & \colhead{$f_D$} & \colhead{$\log x_e$} \\
\colhead{} & \colhead{(mK)} & \colhead{($\mathrm{K \, km \, s^{-1}}$)} & \colhead{($10^{15} \; \mathrm{cm^{-2}}$)} &  \colhead{(K)} & \colhead{($\mathrm{Jy/(34^{\prime\prime})^2}$)} & \colhead{} & \colhead{} } %
\startdata %
\sidehead{G34.43+0.24} \cline{1-3}  
MM1 & 51 & $10.1\pm0.1$ & $4.71\pm0.09$ & 34 & 6.17 & $19\pm4$ & $-4.98$ \\
MM2 & 66 & $23.8\pm0.1$ & $11.2\pm0.2$ & $> 34$\tablenotemark{b} & 4.97 & $<6\pm1$ & $-5.30$ \\
MM3 & 63 & $5.1\pm0.1$ & $2.32\pm0.06$ & 32 & 1.35 & $9\pm2$ & $-5.27$ \\
MM4 & 77 & $13.0\pm0.1$ &  $6.0\pm0.1$ & 32 & 2.74 & $7\pm1$ & $-5.77$ \\
MM5 & 57 & $6.64\pm0.08$ &  $3.0\pm0.1$ & \nodata & 0.92 & $14\pm3$ & $-6.01$ \\
MM6 & 52 & $5.37\pm0.09$ & $2.44\pm0.09$ & \nodata & 0.67 & $13\pm3$ & $-6.62$ \\
MM8 & 55 & $5.2\pm0.1$ & $2.42\pm0.07$ &\nodata & 0.72 & $10\pm2$ & \nodata \\
MM9 & 54 & $4.68\pm0.08$ & $2.12\pm0.07$ & \nodata & 0.75 & $16\pm3$ & $-6.28$ \\
\hline
\sidehead{IRAS~18151-1208} \cline{1-3}  
MM1 & 54 & $16.13\pm0.09$ &  $7.8\pm0.2$ & 27.3 & 2.41 & $6\pm1$ & $-5.37$ \\
MM2 & 59 & $12.0\pm0.1$ &  $5.6\pm0.1$ & 19.4 & 1.85 & $10\pm2$ & $-5.71$ \\
MM3 & 49 & $8.33\pm0.08$ & $3.8\pm0.2$ & \nodata & 0.81 & $8\pm2$ & $-6.33$ \\
\hline 
\sidehead{IRAS~18223-1243} \cline{1-3}
MM1 & 69 & $20.3\pm0.1$ & $9.4\pm0.2$ & 31 & 1.74 & $3.7\pm0.7$ & $-5.80$ \\
MM2 & 60 & $10.10\pm0.09$ & $4.6\pm0.1$ & \nodata & 0.54 & $6\pm1$ & $-6.03$ \\
MM3 & 72 & $8.9\pm0.1$ & $4.1\pm0.1$ & 18 & 0.71 & $7\pm1$ & $-5.77$ \\
MM4 & 73 & $4.6\pm0.1$ & $2.09\pm0.09$ & \nodata & 0.24 & $6\pm1$ & $-5.65$ \\
\enddata %
\tablenotetext{a}{~Dust teperature based on SED fits in literature \markcite{Rathborne2005ApJ630,Marseille2008A&A488,Beuther2010A&A518}(Rathborne {et~al.} 2005; Marseille {et~al.} 2008; Beuther {et~al.} 2010).}
\tablenotetext{b}{~Dust temperature adopted from G34.43-MM1.}
\end{deluxetable} %

\end{document}